# Proposal for an experiment to demonstrate the block universe

Alan McKenzie

*Lately of University Hospitals Bristol NHS Foundation Trust, Bristol UK*

**Abstract**

While the concept of the block universe has a most respectable scientific provenance, many physicists nevertheless do not accept that future events are just as embedded in spacetime as are those of the present and the past. This paper proposes an experiment to demonstrate the block universe using interplanetary distances and high relative speeds such as those accessible through the NASA Mars Reconnaissance Orbiter. The proposed experiment hinges on signals exchanged between Earth and the Orbiter which reveal a time interval of some tens of milliseconds in the Earth's future which is already in the Orbiter's past. Since this experiment can be performed at any time, and since the magnitude of the time interval can in principle be increased in proportion to the distances and speeds over which it is performed, an observer can always be found for whom the past is in another observer's future. The only explanation that fits these observations is a block universe in which all events in the past, present and future of any observer are equally enfolded into spacetime.

1.  **Introduction**

Arguments over the question of free will can be traced back to the Ancient Greeks and doubtlessly stem from the millennia before recorded history. For many, Hermann Minkowski settled the debate with his formulation of spacetime and the resulting *block universe*, in which, for any observer, the future is just as embedded as the past and present. The theory was all the more credible, having "sprung from the soil of experimental physics" as Minkowski himself put it [1]. Sceptics, though, were encouraged by the burgeoning quantum theory, equally grounded in experiment, and which demonstrated the essential randomness of quantum events. So we are apparently left with fundamentally undetermined quantum outcomes in a completely "predetermined" block universe.

Any worthy theory of the universe must unravel this paradox to the satisfaction of both parties. Everett's Many Worlds interpretation (MWI) [2] and its successors (e.g., [3], [4], [5]) go some way to resolving the dilemma, but not entirely: in MWI, while events in a given observer's past and present are fixed as in the block universe, the future nevertheless contains an infinite number of branches. In order to obviate this problem, McKenzie [6] proposed a multiverse of individual, stand-alone block universes populated with quantum outcomes determined at the *multiverse* level by a discrete formulation of the familiar equations of quantum mechanics.

A solution such as this is so radical, however, that any remaining doubts as to the validity of the block universe need to be readdressed. Even with strong support for the block universe from authors in the 1960s such as Rietdijk [7] and Putnam [8], who probably felt they had written the last words on the matter, the old arguments have resurfaced like the Hydra's heads. Even the vivid encapsulation of these ideas in Penrose's so-called Andromeda Paradox [9] did not satisfy a significant number of physicists, who regard Minkowski spacetime as an idealized concept not capable of accounting for the time evolution of real, complex systems (see, for instance, Ellis [10], Sorkin [11]).

In the final analysis, the problem seems to be this: while quantum uncertainty can be demonstrated directly (for instance, by a single-photon, two-slit experiment), it is only the *underlying concepts* of the block universe that are rooted in experiment (such as the null result of the Michelson-Morley experiment): no experiment has yet exposed the block universe directly.

## 2. Ingredients of an experiment

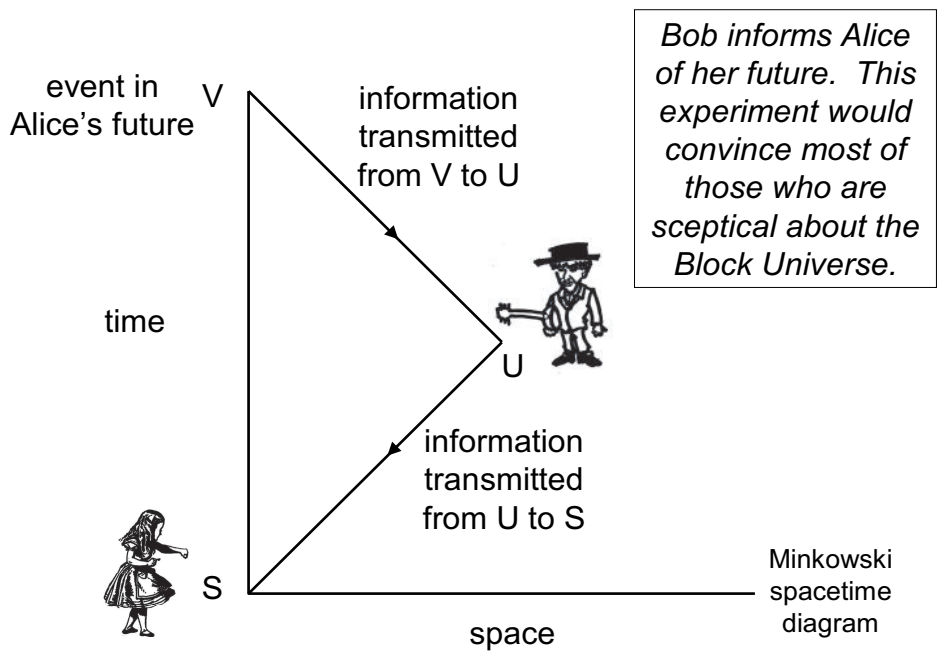

Figure 1: While this experiment might convince all but the most sceptical, it is clearly unrealistic. Nevertheless, it contains the ingredients of a more plausible experiment.

So that is the purpose of this paper – to propose an experiment that will demonstrate the block universe as directly as possible. Figure 1 shows the type of experiment that would be most likely to convince the majority of those who are sceptical about the block universe. If Alice's future is "already written" (strictly a meaningless phrase for a timeless block universe, but it captures the sentiment), then, if the information about her future could somehow be transmitted to Bob at an earlier time, he could, in turn, inform Alice of her future.

Of course, such an experiment is unachievable. Basically, it fails because it would require part of the wave function of the whole universe to evolve unitarily in one "direction" (including Bob's activities to receive and re-send information) while, at the same time, another part of the wave function would have to evolve in the opposite direction (the very propagation of the information about Alice's future).

### 3. Determining simultaneous events

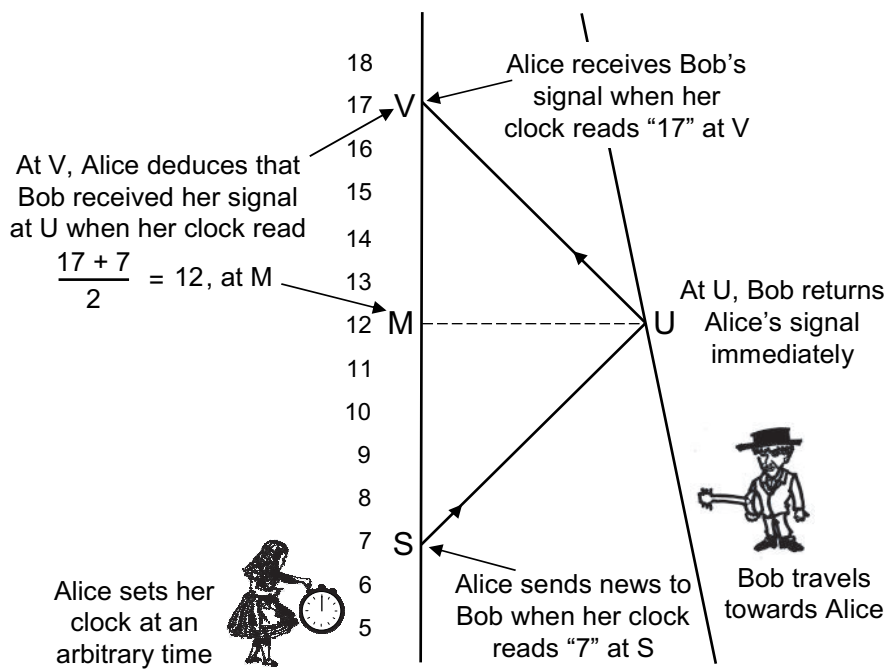

Figure 2: This shows how Alice determines the time that was on her clock at the moment when Bob received her message. It is the time mid-way between the time she transmitted the message and received the acknowledgement.

Figure 1 contains the ingredients of the kind of experiment we are looking for: Alice accepts help from an agent, Bob, who has knowledge about her future. So, while remaining in Alice's reference frame on Earth, let us retain Bob, but have him approach Alice in a spaceship so far away that there are noticeable time delays in conversations between them. Suppose that Alice transmits a radio message to Bob. As shown in Figure 2, Alice starts her clock (which need not be accurate: it is only necessary that it marks regular time intervals) at the moment she transmits her message. According to a prearranged protocol, Bob returns a radio message to Alice at the moment he receives her signal.

From the time she transmits the message (at S) to the time she receives Bob's acknowledgement (at V), she does not know at exactly which moment Bob receives her message (at U) – all she knows is that it is some moment within that time interval. When she receives Bob's acknowledgement, she notes the time on her clock and deduces that Bob received her message when her clock reading was the mean of the times of sending and receiving signals (at M). (This is simply half of the time taken for the return trip.) So, in Alice's reference frame, M and U are simultaneous. Notice that, because the speed of light is constant, Bob's motion relative to Alice does not affect that fact that the time taken for Alice's radio message to reach Bob is the same as that taken for the return signal to travel from Bob back to Alice.

## 4. Structure of the experiment

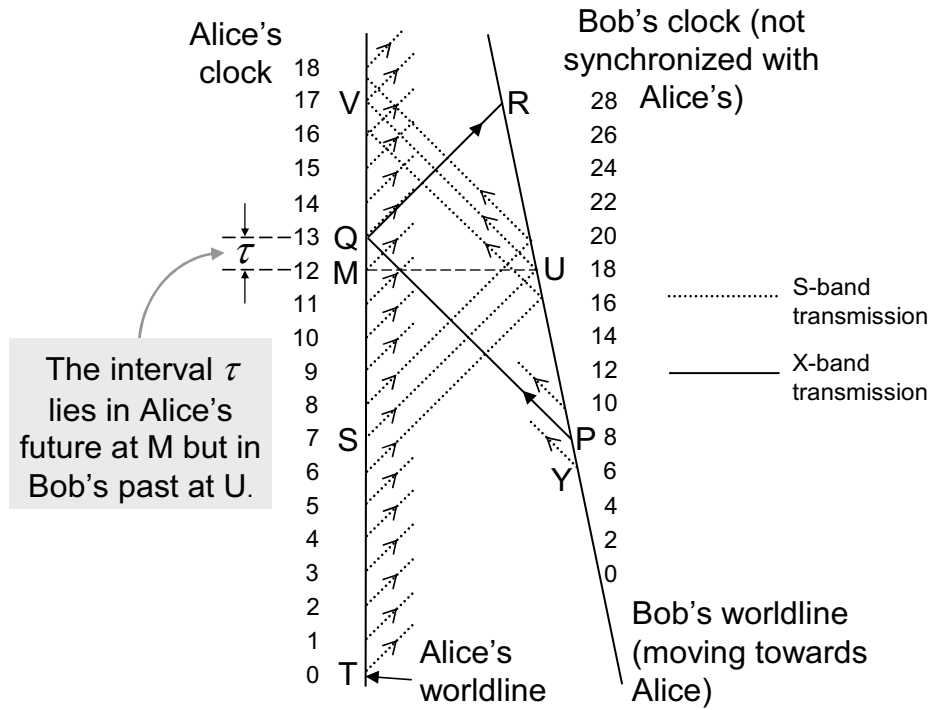

Figure 3: Alice sends a series of signals to Bob, each containing her clock reading, using S-band microwaves. Bob returns these signals immediately, adding to each returned signal his own clock reading as well as the clock reading received from Alice. He also sends a single signal to Alice containing his clock reading using X-band microwaves, and Alice returns it to Bob. The returned signal contains Alice's clock reading together with the clock reading received from Bob.

Since Bob is moving towards Alice, the events M and U which are simultaneous in Alice's reference frame cannot be simultaneous in his reference frame. It is this asymmetry that is at the root of this paper's proposed experiment to demonstrate the block universe. Figure 3 builds upon the experiment in Figure 2, with Alice this time sending not a single message but a series of signals to Bob using microwaves in the S-band region. Each signal contains the reading on her clock at the moment when the signal is sent. The series begins at T in the diagram, when she starts her clock.

Bob receives Alice's first transmission at Y in the diagram, which is some time after he starts his own clock. He has a transponder which sends signals back to Alice with a delay of less than one microsecond using the same S-band frequency. Each signal that Bob sends back to Alice contains the original clock reading that she sent him together with the reading from his own clock at the moment he replies. Bob's clock is not synchronized with Alice's, and, in any case, it runs at a different rate. (It runs at a different rate simply to emphasize that the experiment does not rely on Alice's and Bob's clocks being synchronized, and that time dilation, which is in any case small, does not affect the outcome.)

| Label | Alice's clock | Event in Alice's frame of reference | Label | Bob's clock | Event in Bob's frame of reference |
|---|---|---|---|---|---|
| T | 0 | Begin series of S-band transmissions to Bob, each containing Alice's current clock reading. | Y | 6 | Receive first of series of S-band transmissions each containing Alice's clock-reading. S-band transponder returns same data to Alice plus Bob's current clock reading. |
| Q | 13 | Receive X-band signal containing Bob's clock reading of "8" at P. Alice's X-band transponder returns same data to Bob at R plus Alice's current clock reading of "13" | P | 8 | Transmit signal on X-band containing Bob's current clock reading of "8" |
| V | 17 | Receive transponded S-band signal containing Alice's clock reading of "7" at S plus Bob's clock-reading of "18" at U | R | 28 | Receive transponded X-band signal from Alice containing Bob's clock reading of "8" at P plus Alice's clock reading of "13" at Q |
| M | 12 | Clock-reading at S ("7") and V ("17") mean that signal reached Bob at U (when Bob's clock read "18") when Alice's clock read (7+17)/2 = 12. | U | 18 | Clock reading at P ("8") and R ("28") mean that signal reached Alice at Q (when Alice's clock read "13") when Bob's clock read (8+28)/2 = 18. |

Table 1

At a time of his own choosing (P in the diagram), in addition to responding to Alice's S-band transmissions, Bob sends a single signal to Alice containing his current clock reading. This single transmission is in the X-band microwave frequency so that it may be distinguished from the other signals in the S-band. (The small Doppler shift between transmission and receipt of the signals in these experiments has no bearing on the outcome of the experiment.) Alice receives this X-band signal at Q. She has a transponder which returns a signal to Bob, again at the X-band frequency, again with a delay of less than one microsecond. This returned signal contains the clock reading that Bob sent her from P together with her own clock reading at Q. Bob receives her returned signal at R. This sequence of events is summarized in Table 1.

## 5. Analysing the experiment

This completes the experiment. It is in the analysis that the block universe is revealed. After the experiment, Alice and Bob exchange the data that they have each gathered. From Bob's clock reading at the moment the solitary X-band signal returned to him at R, together with his clock reading at P when he first sent it out, they calculate Bob's clock reading at U, when, in Bob's reference frame, the X-band signal reached Alice at Q. In Figure 3, this clock reading at U is "18". This mirrors the calculation of the return-trip time that Alice applied to her returned signal in Figure 2.

Now Alice and Bob examine the records of the returned S-band signals which Alice sent to Bob and received back from him through his transponder. For each returned signal, they calculate Bob's clock reading at the moment that it reached him, again by taking half of the return-trip time. They are particularly interested in the S-band signal from Alice that reached Bob when his clock reading was "18" at U. In this example, this turns out to be the signal sent by Alice when her clock reading was "7" at S and transponded back to her, reaching her when her clock reading was "17" at V. The return-trip time from these two clock readings means that Alice's clock reading was "12" at M when, in her reference frame, her S-band signal reached Bob at U.

In summary, in Alice's reference frame, an event in her worldline at M was simultaneous with an event in Bob's worldline at U, while this same event, U, in Bob's reference frame was simultaneous with an event in Alice's worldline at Q, which, from M, is in Alice's *future*. Let $\tau$ denote the interval from M to Q.

In other words, at the event U in Bob's reference frame, the interval $\tau$ is already in his past and so cannot be altered. Therefore, at the moment M in Alice's reference frame, her future up to an interval $\tau$ ahead was already fixed.

## 6. Minkowski geometry of the experiment

The root cause of this phenomenon, the constancy of the speed of radio waves for all observers, is illustrated in Figure 4, in which a section of Figure 3 is redrawn and enlarged. The worldlines of signals travelling at the speed of light in any rest frame are drawn at an angle of 45° in a Minkowski diagram. This is because distances drawn in the diagram in the vertical (time) direction are the same as the equivalent ones in the horizontal (space) direction divided by $c$. The geometry of Figure 4 is meant to be self-explanatory, and illustrates the well known result of special relativity that the time interval $\tau$ is given by $\tau = \dfrac{xv}{c^2}$ where $x$ is the distance MU between Alice and Bob and $v$ is the speed at which Bob approaches Alice.

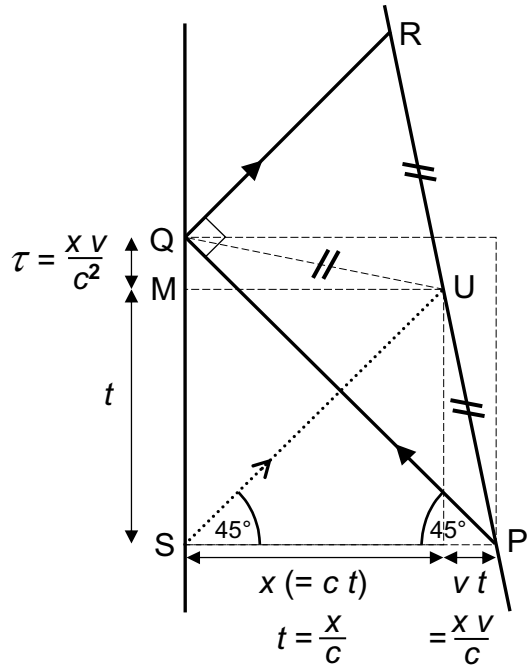

Figure 4: Geometrical derivation of the time interval $\tau$. Note the two squares constructed by virtue of the two diagonals at 45°, namely SU and PQ. This guarantees that PU and QU are equal. PU and RU are also equal because U has to be at the mid-point of the return trip PQR. The length $vt$ is the distance Bob travels towards Alice at speed $v$ in time $t$, and is equal to QM (after dividing by $c$ because of the way the Minkowski diagram is drawn).

In devising an experiment to reveal $\tau$, it is important to use a transponder that returns a signal in the shortest time practicable, which, for microwave technology, is typically of the order of one microsecond, although allowance can be made for this known delay. It is cleaner if $\tau$ is significantly larger than the transponder delay: the less the interpretation required for the experiment, the greater its impact. (The intensities of light and microwave signals returned without a delay from corner-cube retroreflectors are too low to carry clock-reading information fast enough.)

This suggests that $\tau$ should be of the order 100 microseconds or greater. For realistically achievable relative speeds – in the order of 10 km s$^{-1}$ for space equipment – this implies distances in the order of $10^6$ km, which rules out, for instance, Earth-based satellites such as the International Space Station as it comes into view above the Earth's horizon.

## 7. An experiment on an interplanetary scale

The stylized orbits of Earth and Mars in Figure 5 show that interplanetary distances meet the requirements. When the Earth is approaching Mars tangentially, it is closing in on it at a relative speed of 14.2 km s$^{-1}$ at a distance of $172 \times 10^6$ km. This produces a value of 27 ms for $\tau$, much greater than the transponder delay.

The NASA Mars Reconnaissance Orbiter (MRO) offers one possible route for implementing the experiment, with a data-handling capability of several Mbs [12]. In order to demonstrate the interval $\tau$ clearly, the repetition rate of S-band signals sent to the satellite should ideally be in the order of 1 kHz or greater, so long as that leaves enough "room" in each pulse to contain the clock readings.

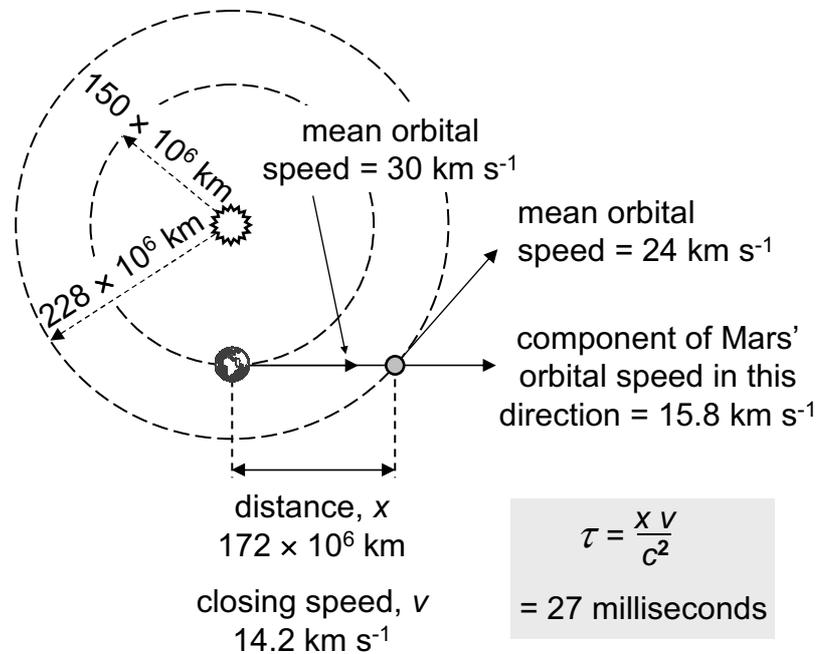

Figure 5: The orbits of Earth and Mars are stylized, and the orbital radii and speeds are means.

The outcome of the experiment is in the signals received at V and R (referring to Figure 3). The signal received on Earth at V would state, effectively, that the moment, M, on Earth was the moment, U, on the MRO when it received a signal transmitted from Earth nearly ten minutes earlier. The signal received on the MRO at R would effectively state that the same moment, U, on the MRO coincided with the moment, Q, on Earth, 27 milliseconds later than the moment, M, on Earth, and that the 27 milliseconds prior to Q were already in the past from the standpoint of the MRO.

Needless to say, the same experiment will show that, in the reference frame of the MRO, the time interval 27 milliseconds into its future is already in the past in the reference frame of the approaching Earth. Such is relativity.

## 8. Discussion

Those who question the idea of the block universe will be unimpressed with this result. In addition to the above-cited authors who accept the results of special relativity but dispute their interpretation, Clifton and Hogarth [13] speak for those such as Stein [14], [15] who directly challenged the arguments of Rietdijk and Putnam (op. cit.) and similar work by other authors. In the context of the proposed experiment, the issue that will underlie all of the objections to its block-universe interpretation is that, while an event, U, on Mars may be simultaneous with an event, M, on Earth, U will not be part of the Earth's history of events until nearly ten minutes later. Until that time, on Earth, heuristically speaking, "anything may happen". Equally, an observer on the MRO could say the same thing, and these views may be summarized in Figure 6 taken from [6].

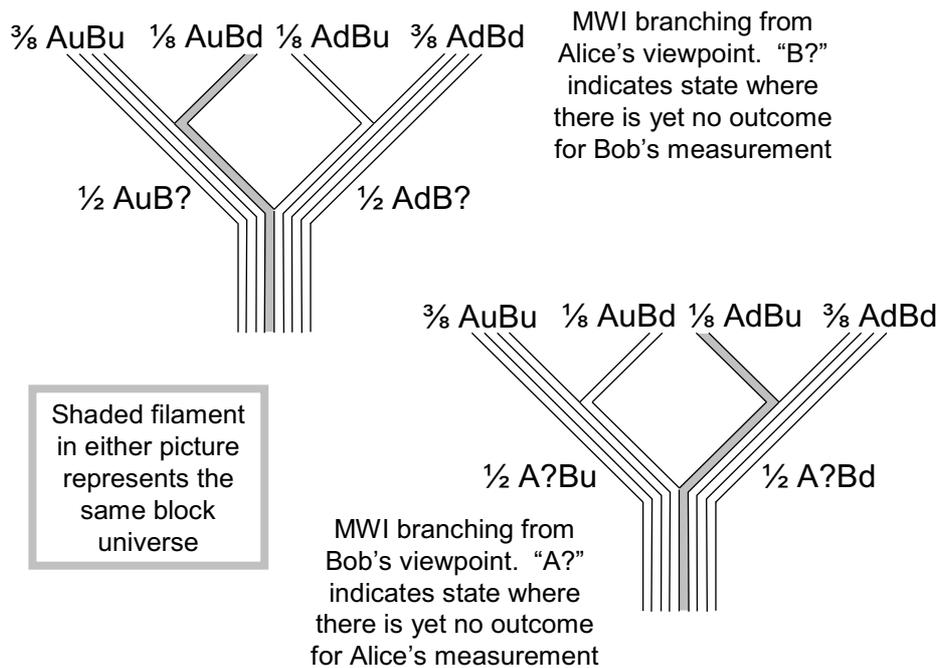

Figure 6: Two Many-Worlds interpretations (MWI) of the same quantum-entanglement experiment from the different perspectives of Alice on Earth (top left) and Bob on the Mars Reconnaissance Orbiter (bottom right). The trunk and branches in each MWI tree are split into filaments within each of which the outcome of the experiment is certain (reproduced from [6]).

The diagram is drawn from two perspectives: that of Alice, whom we shall assume is on Earth and that of Bob, whom we shall place on the MRO. Their experiences are represented by the two MWI "trees". The lettering, ⅛AuBd etc., records the outcomes of a quantum-entanglement electron-spin experiment with spin filters orientated at a mutual angle of 120°, and where "A", "B", "u" and "d" stand for "Alice", "Bob", "spin-up" and "spin-down" respectively. The fractions ⅜ and ⅛ are the probabilities of an outcome such as that of Alice observing an up-spin on her electron and Bob observing a down-spin on his (⅛AuBd).

Critics of the block-universe interpretation regard only the trunk of the MWI tree as in the past and fixed. Above the trunk, it is fluid. So, taking Alice's viewpoint on Earth in the top-left tree, Alice has observed either "u" or "d" for her electron in the two main left and right branches respectively, but she does not yet know, in either of those branches, the outcome of Bob's measurement on the MRO. He will have detected the spin of his own electron some minutes earlier, but, until the news reaches Alice, her uncertainty is signified by the question mark, "?", next to "B" in both of the branches.

It is only when news finally reaches Alice that the outcome of Bob's experiment also becomes part of Alice's past. Until that moment, say the critics, the future is uncertain for Alice and, equally, it is uncertain for Bob until news of Alice's outcome reaches him, as depicted in the MWI tree at the bottom right of the diagram.

However, these uncertainties are not absolute: they are relative to Alice's and Bob's individual perspectives. During the time that Alice is uncertain of Bob's outcome, he is not: for him, his outcome is already history. The MWI trees and branches in Figure 6 have been sectioned into eight filaments so that the combined thicknesses of the branches represent the probabilities of the final outcomes. The two grey-highlighted filaments that culminate in ⅛AuBd at the top of the two MWI trees, while accommodating in one case Alice's uncertainty and Bob's in the other, are nevertheless identical: the outcome of the quantum-entanglement experiment in that filament is certain, and guaranteed to be ⅛AuBd. The personal uncertainties of Alice and Bob arise from not knowing which particular filament that particular version of them is in. Each filament, of course, is a block universe (see [6] for further discussion).

## 9.  Conclusion

Of course, there is nothing special about the time and date when the experiment is performed. As long as planetary orbits allow, it can be carried out at any time with the same result. Nor is any significance attached to the Earth-Mars location of the experiment. The separation could be billions of light-years in principle[1], although its outcome would have to be taken on trust by extrapolating from experiments which can be performed within an accessible time. An observer can always be found for whom the past is in another observer's future. The only explanation that fits these observations is a block universe in which all events in spacetime are embedded.

Certainly, not all of those who are sceptical of the theory of the block universe will be persuaded by its demonstration in the satellite experiment, and fewer still will be moved simply by the description of the experiment in this paper. That is all to the good: it is necessary to be critical of such an extraordinary concept as the block universe, and the theory will be the more robust if it survives such scrutiny, convincing its former detractors to share the sentiments of Einstein who, in a letter of condolence to the family of Michele Besso, his close friend, famously wrote [16]:

> *Now Besso has departed from this strange world a little ahead of me. That means nothing. People like us, who believe in physics, know that the distinction between past, present and future is only a stubbornly persistent illusion.*

---

[1] In such a thought experiment, care must be taken not to include the general relativistic expansion of space in the velocity component when calculating $\tau$. It is only the "local" speed that affects $\tau$.